\documentclass[12pt,aps,prb,preprint]{revtex4}
\usepackage{amsmath}    
\usepackage{graphicx}   
\usepackage{bm}

\begin{document}

\title{Relativistic momentum and kinetic energy, and $E=mc^2$}

\author{Ben Yu-Kuang Hu}
\altaffiliation[On sabbatical leave at ]{Department of Physics, University of Maryland, College Park, MD 20742-4111.}
\email{yhu@umd.edu}
\affiliation{Department of Physics,
University of Akron, Akron, OH~44325-4001.}

\date{\today}

\begin{abstract}
Based on relativistic velocity addition and the conservation of
momentum and energy, I present derivations of the expressions for the
relativistic momentum and kinetic energy, and $E=mc^2$.
\end{abstract}
\maketitle
\newpage

\section{Introduction}

The standard formal way that expressions for the relativistic momentum $\bm p$ and the relativistic kinetic energy $T$, and the mass--energy relationship
$E=mc^2$ are derived in upper-level undergraduate textbooks\cite{textbooks} is by first introducing Lorentz transformations and 4-vectors,
and then defining the 4-momentum vector $p^\mu = m\, dx^\mu/d\tau$ ($\mu = 0, 1, 2, 3$), where $\tau$ is the proper time.  It is then postulated, backed up
by extensive experimental observations, that in an isolated system all the components of $p^\mu$ are conserved.
The spatial components of $p^\mu$ reduce to $m{\bm v}$ in the non-relativistic limit, and hence correspond
to the components of the relativistic momentum.   The temporal component reduces in the non-relativistic limit
to $mc^2 + mv^2/2$, and therefore is identified as the total energy, composed of the rest-mass and kinetic energies.
Logically, there is of course nothing wrong with this approach.
Pedagogically, however, it is probably helpful to have more intuitive derivations.  To this end,
through the years many have been published.\cite{derivations}

This paper describes relatively simple and concise derivations of the relativistic forms of ${\bm p}$
and $T$, and $E=mc^2$, based on (i) conservation of momentum and energy in the collisions of two particles, and
(ii) the velocity addition rules.  Momentum and energy conservation should be familiar concepts to students, and the
velocity addition rules can be quite simply derived from the constancy of the speed of light in all inertial reference
frames.\cite{Constant_c}
In each derivation, collisions are viewed in the center-of-momentum frame of reference, $S_{\rm cm}$, in which both particles have momenta that are equal
in magnitude and opposite in direction, and in the laboratory frame of reference, $S_{\rm lab}$,
in which one of the particles initially is at rest.  Imposition of the conservation laws gives the desired expressions.

To simplify the algebra, velocities in this paper are expressed in units of $c$, the speed of light.
Hence velocities are dimensionless, and $c=1$. To obtain the
standard dimensional expressions, replace all velocities in the expressions given here by $v\rightarrow v/c$,
and multiply all masses by $c^2$ in order to obtain
energy.  Also, in this paper primes on variables denote ``after collision."

\section{The derivations}

First, let us recall the relativistic velocity transformation rules. Let
$\tilde{S}$ be an inertial frame moving with velocity $(u,0)$
with respect to frame $S$.  If a particle has velocity $(v_x,v_y)$
in frame $S$, the components of its velocity in frame $\tilde{S}$
are\cite{textbooks,Constant_c}
\begin{subequations}
\label{Eq:velocity-transform}
\begin{eqnarray}
\tilde{v}_x &=& \frac{v_x - u}{1 - v_x u}, \\
\tilde{v}_y &=& \frac{v_y\sqrt{1-u^2}}{1-v_x u}.
\end{eqnarray}
\end{subequations}

\subsection{Relativistic momentum}
\label{sec:momentum}

From dimensional analysis and the vector\cite{vector-scalar} nature of momentum,
the momentum of a particle of mass $m$ travelling with velocity
${\bm v}$ must have the form \begin{equation} {\bf p} =
m\gamma_v\,\bm v,\label{Eq:form-of-p}
\end{equation}
where $\gamma_v$ is an unknown function to be determined.  Since ${\bf p} = m{\bm v}$ for
non-relativistic velocities, $\gamma_{v\rightarrow 0} = 1$.

Consider the case where the particles are identical, hence $m_1 =
m_2 = m$.  Let the motion of the particles be in the $x$--$y$
plane and their initial velocities in $S_{\rm cm}$ be $\pm(v,0)$.  Assume that the particles barely graze each other,
so that in the collision each particle picks up a very small $y$-component of the velocity
of magnitude $\delta v$ in $S_{\rm cm}$.
[See Fig. 1(a).]  Their speeds in
$S_{\rm cm}$ do not change because the collision is elastic, and
hence their velocities after the collision are $\pm(\sqrt{v^2 -
(\delta v)^2}, \delta v) = \pm(v\sqrt{1 - (\delta v/v)^2},\delta v) \approx
\pm(v,\delta v)$, to first order in $\delta v$.  Because $\delta v$ is assumed to be very small, we ignore all
terms of order $(\delta v\!)^2$ and
higher.

Now consider the collision in the laboratory frame of reference
$S_{\rm lab}$ that is moving with velocity $(-v,0)$ with respect
to $S_{\rm cm}$. [See Fig. 1(b).]  The pre-collision velocities of the particles
$S_{\rm lab}$, using Eqs.~(\ref{Eq:velocity-transform}) on the $S_{\rm cm}$ velocities $\pm(v,0)$, are
${\bm v}_{\rm 1,lab} = (w,0)$, where
\begin{equation}
w = \frac{2v}{1 + v^2},
\label{Eq:w}
\end{equation}
and ${\bm v}_{\rm 2,lab} = (0,0)$.
After the collision,
transforming the post-collision $S_{\rm cm}$ velocities
$\pm(v,\delta v)$ to the $S_{\rm lab}$ frame we obtain, to first order in $\delta v$,
\begin{subequations}
\label{Eq:py}
\begin{eqnarray}
{\bm v}_{\rm 1,lab}' &\approx& \left(w, \frac{\delta
v\,\sqrt{1-v^2}}{1+v^2}\right),
\label{Eq:p1y}\\
{\bm v}_{\rm 2,lab}' &\approx& \left(0,-\frac{\delta
v\,\sqrt{1-v^2}}{1-v^2}\right).\label{Eq:p2y}
\end{eqnarray}
\end{subequations}

The $y$-component of the total momentum before the collision
is zero, and hence by conservation of momentum, after the collision
\begin{equation} \left({\bm p}_{{\rm 1,lab}}' + {\bm p}_{{\rm 2,lab}}'\right)_{y} = m\, \gamma_{|{\bm v}_{{\rm 1,lab}}'|}\;
v_{{\rm 1,lab},y}' + m \,\gamma_{|{\bm v}_{{\rm 2,lab}}'|}\;v_{{\rm
2,lab},y}' = 0.\label{Eq:pytot}\end{equation}
Since $v_{{\rm lab},y}'$ terms are of order $\delta v$ and we are ignoring terms of order $(\delta
v\!)^2$, it is sufficient in Eq.~(\ref{Eq:pytot}) to evaluate $|{\bm v}_{\rm 1,lab}'|$ and $|{\bm v}_{\rm 2,lab}'|$ to zeroth order in
$\delta v$ ({\em i.e.}, ignoring $\delta v$ altogether). Substituting $|{\bm v}_{\rm 1,lab}'| \approx w$, $|{\bm
v}_{\rm 2,lab}'| \approx 0$, and the $y$-components of the
velocities from Eqs.~(\ref{Eq:p1y}) and (\ref{Eq:p2y}) into Eq.~(\ref{Eq:pytot}) and using $\gamma_{v'\rightarrow 0} = 1$
yields
\begin{equation}
\frac{\gamma_w}{1+v^2} -\frac{1}{1-v^2}=0.
\label{Eq:pytotal}
\end{equation}
This, together with Eq.~(\ref{Eq:w}) gives
\begin{equation}
\gamma_w = \frac{1+v^2}{1-v^2} = \left(1 -
\left[\frac{2v}{1 + v^2}\right]^2\right)^{-1/2} =
\left(1-w^2\right)^{-1/2}\ .
\label{eq:gamma_desired_result}
\end{equation}

\subsection{Relativistic kinetic energy}
\label{sec:kinetic}

Dimensional analysis and the scalar\cite{vector-scalar} property of kinetic energy imply that its form is
\begin{equation}
T = m\, G\!(v),\label{eq:assume_E}
\end{equation}
where $m$ is the mass of the particle, $v = |{\bm v}|$ is its speed and
the function $G(v)$ is to be determined.

Consider an elastic head-on collision
between two particles, of mass $m$ and $M \gg m$, with speeds in $S_{\rm cm}$ of $v$ and $V$, respectively.
In $S_{\rm cm}$, the particles simply reverse
directions, and the motion is one-dimensional. [See Fig. 2(a).]
Assume that the mass $M$ is so large that in frame $S_{\rm cm}$ its speed $V \ll 1$, and hence we can use
the non-relativistic expressions for the momentum and kinetic energy of mass $M$.
The magnitudes of the momenta of $m$ and $M$ are equal in $S_{\rm cm}$, implying
\begin{equation}
m\gamma_v v = MV. \label{Eq:cons_mom}
\end{equation}

The $S_{\rm cm}$ frame pre- and post-collision velocities of mass $m$
are ${\bm v}_{\rm cm} = v$  and ${\bm v}_{\rm cm}' = -v$ respectively, and of mass $M$ are
${\bm V}_{\rm cm} = V$ and ${\bm V}'_{\rm cm} = -V$, respectively.
Transforming these to the $S_{\rm lab}$ frame, which is moving at velocity $-V$ with respect to $S_{\rm cm}$ [see
Fig. 2(b)],
gives ${\bm v}_{\rm lab} = (v + V)/(1 + v V)$, ${\bm v}_{\rm lab}' = (-v + V)/(1-vV)$, ${\bm V}_{\rm lab} = 0$ and
${\bm V}_{\rm lab}' = 2 V/(1 + V^2)$.  By conservation of kinetic energy in an elastic
collision in the $S_{\rm lab}$ frame and Eq.~(\ref{eq:assume_E}),
\begin{equation}m\, G(|{\bm v}_{\rm lab}|) = m\, G(|{\bm v}_{\rm lab}'|) +
\frac{M}{2}|{\bm V}_{\rm lab}'|^2.\label{Eq:cons_energy}\end{equation}
Expanding
$|{\bm v}_{\rm lab}|$, $|{\bm v}_{\rm lab}'|$ and $|{\bm V}_{\rm lab}'|$ to first order in $V$,
\begin{subequations}
\begin{eqnarray}
|{\bm v}_{\rm lab}| &\approx&
(v + V) (1 - v V) \approx v + V(1-v^2),\\
|{\bm v}_{\rm lab}'| &\approx&
(v-V) (1 + vV) \approx v - V (1-v^2),\\
|{\bm V}_{\rm lab}'| &\approx& 2 V (1 - V^2) \approx 2 V,
\end{eqnarray}
\label{Eq:expansions}
\end{subequations}
and substituting these into the Taylor expansions of the $G$ terms about $v$ in
Eq.~(\ref{Eq:cons_energy}) gives, to first order in $V$,\cite{term_m2v2squared}
\begin{equation}
m \left(G(v) + \left[\frac{d  G(u)}{d
u}\right]_{v}\!V(1-v^2)\right) = m \left(G(v) -
\left[\frac{d  G(u)}{d
u}\right]_{v}\!V(1-v^2)\right) + 2 MV^2.
\label{Eq:cons_mom_orderV}
\end{equation}
Substituting $2M V^2 = 2 m \gamma_v vV$
[from Eq.~(\ref{Eq:cons_mom})] into Eq.~(\ref{Eq:cons_mom_orderV}) leads to
\begin{equation}
\left[\frac{dG}{du}\right]_{v} = \frac{\gamma_v\, v}{1 -
v^2} = \frac{v}{(1 - v)^{3/2}},
\end{equation}
which upon integration yields
\begin{equation}
G(v) - G(0) = \left[\frac{1}{(1-u^2)^{1/2}}\right]_{u=0}^{u=v}
 = \gamma_v - 1.\label{eq:18}  \end{equation}
 Since the kinetic energy vanishes when $v$ is zero, $G(0) = 0$, and hence
 Eqs.~(\ref{eq:assume_E}) and (\ref{eq:18}) imply that (reintroducing $c$)
$T = m\bigl(\gamma_v -1\bigr)c^2$.

\subsection{$E=mc^2$}
\label{sec:E=mc2}

Consider the initial situation as in Sec. \ref{sec:kinetic}, except that the speed $V$ of mass $M$ can be
relativistic, and after collision the two particles merge into one composite particle.  In $S_{\rm cm}$,
$M\gamma_V V = m\gamma_v v$, and after the collision the composite particle is stationary.
In $S_{\rm lab}$ which is moving with velocity $-V$ with respect to
$S_{\rm cm}$, before the collision particle $M$ is stationary and particle $m$ moves with
velocity $\bm v_{\rm lab} = (v+V)/(1+vV)$, and after the collision the composite particle
moves with velocity $V$.

The total momentum in $S_{\rm lab}$ before the collision is
$\bm P_{\rm lab} = m \gamma_{|\bm v_{\rm lab}|} \bm v_{\rm lab} = m
\gamma_v\gamma_V (u+V)$. If the mass of the composite particle
does not change, then the momentum of the composite particle after
the collision in  $S_{\rm lab}$ would be $(M + m) \gamma_V V
\ne \bm P_{\rm lab}$ in general, violating conservation
of momentum. Therefore, the mass of the composite
particle must change by $\Delta m$ such that momentum is conserved in $S_{\rm lab}$; {\it i.e.},
\begin{equation}
m\gamma_v\gamma_V (v+V) = (M+m+\Delta m)\gamma_V V.
\end{equation}
Substituting $m \gamma_v v = M\gamma_V V$ on the left hand side
and cancelling $\gamma_V V$ on both sides gives
\begin{equation}
\Delta m = m(\gamma_v- 1) + M(\gamma_V - 1).\label{eq:delta_m}
\end{equation}
From Sec.~\ref{sec:kinetic}, the right hand side of
Eq.~(\ref{eq:delta_m}) is equal to $-\Delta T$, the total change
in kinetic energy in $S_{\rm cm}$  (since particles $m$ and $M$
start with speeds $v$ and $V$, respectively, and both are
stationary at the end).  By conservation of total energy, $\Delta
E + \Delta T = 0$, where $\Delta E$ is the energy associated with
the change in mass. Hence, $\Delta E = -\Delta T = \Delta m$ or
(reintroducing $c$, and making the plausible assumption that a
zero mass object with zero velocity has zero energy) $E = mc^2$.
 Finally, combining the results of Sections \ref{sec:kinetic} and
\ref{sec:E=mc2} gives the total energy of a particle of mass $m$
moving with speed $v$, $E + T =E_{\rm total} =  m\gamma_vc^2$.

\section{Concluding remarks}

It should be noted that these derivations do not guarantee that the momentum and total energy are conserved in all
inertial reference frames or in all collisions.  They only show the forms that the momentum, kinetic energy and energy--mass
relation must have, given momentum and energy conservation.   Once these expressions are known,
when the 4-momentum is introduced its components will be recognized as the total energy and momentum.
The covariance of the momentum 4-vector can then be used to demonstrate momentum and
total energy conservation in all inertial frames.  Conservation of momentum can be shown to be a consequence of
conservation of energy,\cite{cons_energy} and, as befitting an experimental science, the conservation of energy ultimately
depends on experimental observations.

\newpage
\begin{center}
\large\underline{Figures}
\end{center}

\begin{figure}[ht]
\includegraphics[scale=0.7]{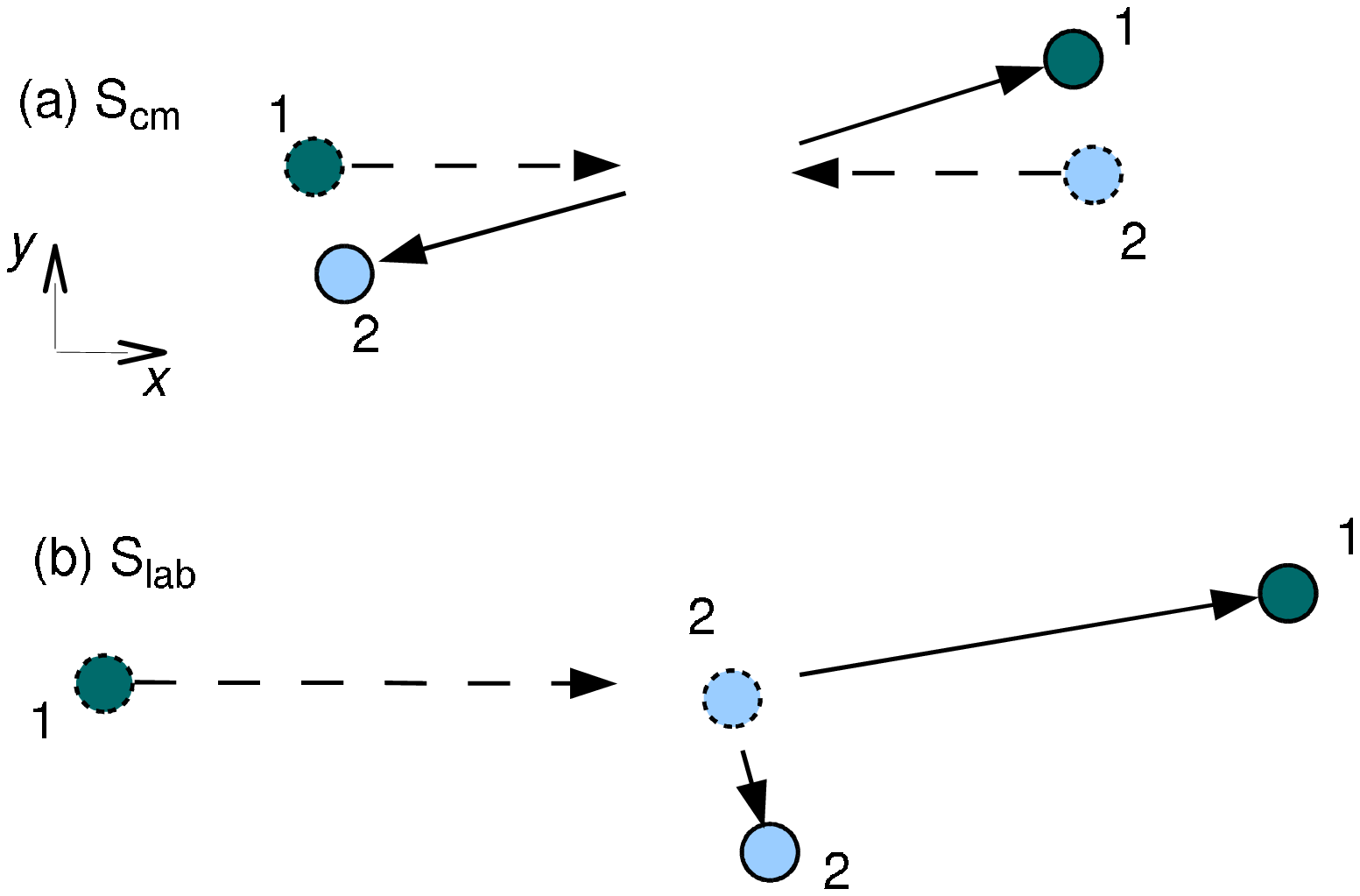}
\caption{Grazing collision between two particles of equal mass, in (a) center-of-momentum and (b) laboratory frames of
reference.  Dashed and solid lines indicate before and after the collision, respectively.}
\end{figure}
\bigskip

\begin{figure}[ht]
\includegraphics[scale=0.7]{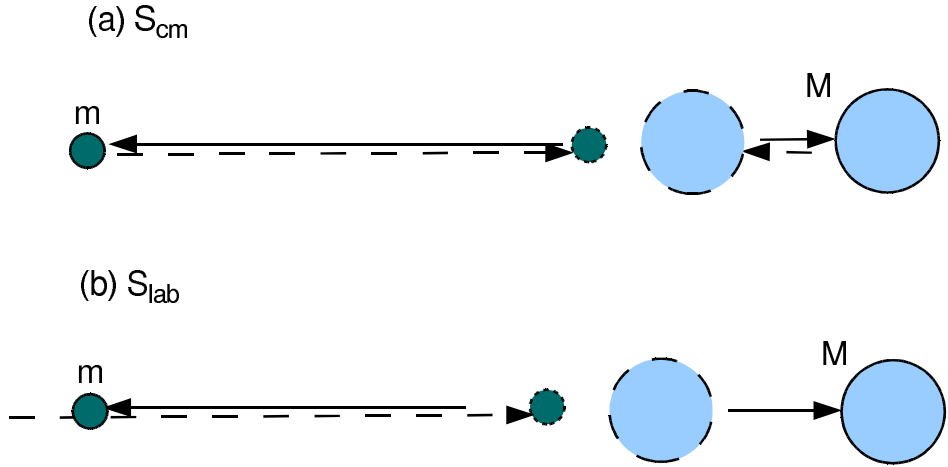}
\caption{Head-on collision between particles of mass $m$ and $M \gg m$, in (a) center-of-momentum and (b) laboratory frames of
reference.  Dashed and solid lines indicate before and after the collision, respectively.}
\end{figure}

\end{document}